\documentclass[twocolumn,showpacs,preprintnumbers,nofootinbib,amsmath,amssymb]{revtex4}

\usepackage{graphicx}
\usepackage{dcolumn}
\input epsf
\def\be{\begin{equation}}
\def\ee{\end{equation}}
\def\bea{\begin{eqnarray}}
\def\eea{\end{eqnarray}}
\def\der{\partial}
\def\d3bar{$\overline{\rm D3}$}
\def\ds{${\rm D7}$}
\def\dt{${\rm D3}$}
\def\eps{\epsilon}
\def\nn{\nonumber}
\begin{document}
\parskip 4pt
\vskip 3cm

\

\preprint{SU-ITP-07/24}
\title {\Large\bf Accidental Inflation in String Theory}
 \author{\bf Andrei Linde and Alexander Westphal \vspace*{0.45cm}
}
\affiliation{ {Department
  of Physics, Stanford University, Stanford, CA 94305-4060,
USA}    }\date{December 10, 2007}
 {\begin{abstract} We show that inflation in type IIB string theory driven by the volume modulus can be realized in the context of the racetrack-based Kallosh-Linde model (KL) of moduli stabilization. Inflation here arises through the volume modulus slow-rolling down from a flat hill-top or inflection point of the scalar potential. This situation can be quite generic in the landscape, where by uplifting one of the two adjacent minima one can turn the barrier either to a flat saddle point or to an inflection point supporting eternal inflation. The resulting spectral index is tunable in the range of $0.93 \lesssim n_s \lesssim 1$, and there is only negligible production of primordial gravitational waves $r<10^{-6}$. The   flatness of the potential in this scenario requires fine-tuning, which may be justified taking into account the exponential reward by volume factors preferring the regions of the universe with the maximal amount of slow-roll inflation. This consideration leads to   a  tentative  prediction of the spectral index $n_s\approx 0.95$ or $n_s\approx 0.93$ depending on whether the potential has a symmetry $\varphi\to-\varphi$ or not.
\end{abstract}}
 \maketitle
\section{Introduction}

In the beginning of  the development of inflationary cosmology our main goal was to find some simple semi-realistic versions of inflationary theory, where inflation could be naturally realized for a wide variety of initial conditions. The search for such models has lead us to the discovery of chaotic inflation \cite{Chaot}, where inflation can be achieved in the simplest models, such as the models of the fields minimally coupled to gravity, with polynomial potentials. Inflation occurs in such models for a wide variety of initial conditions \cite{Linde:1985ub} and may continue eternally \cite{Vilenkin:1983xq,Linde:1986fd}.

In this class of theories the constant part of the scalar field does not have any independent physical meaning; its value enters the theory only because it makes other fields massive, and because it has a potential energy density $V(\varphi)$. As a result, these theories have a weakly broken shift symmetry $\varphi \to \varphi+c$. This is broken only by the $\varphi$-dependence of the potential $V$ and of the masses of particles interacting with the field $\varphi$. It is this property which allows the existence of chaotic inflation potentials such as $m^{2}\varphi^{2}$, which remain flat even for $\varphi \gg M_{p}$ \cite{Chaot}. Shift symmetries forbid the dangerous quantum corrections to the effective potential $\sim M_p^{4} \bigl(\frac{\varphi}{M_{p}}\bigr)^{n}$ often discussed in the literature which renders inflation in these theories easier to achieve;  see \cite{Linde:2007fr} for a recent discussion of this issue. Another popular example of shift symmetry is the symmetry $\theta \to \theta + c$ of the axion potential; this potential appears only after the shift symmetry becomes broken by non-perturbative effects. This symmetry was the basis for the so-called natural inflation scenario \cite{Freese:1990rb}.

However, in supergravity and string theory the scalar fields usually have physical or geometric meaning. For example, in the simplest supergravity models the K\"ahler potential $K = \frac{\Phi\bar\Phi}{M_p^{2}}$ describes K\"ahler geometry. The F-term scalar potential is proportional to $e^{K} \sim e^{\Phi\bar\Phi/ M_p^{2}}$. As a result, the mass of the inflaton field typically acquires a contribution $O(H^{2})$, which tends to prevent inflation. Quite often, one can fine tune the parameters of the theory and achieve desirable flatness of the potential for some small range of the values of the scalar fields. In such models, inflation in a certain sense happens by accident: One must have the parameters of the models fixed in a very narrow range, and in addition, one should hope that the scalar field from the very beginning by some happy accident was in the required range of its values where inflation is possible. A similar situation, which can be called `accidental inflation,' often occurs in string theory.  

An interesting and quite instructive model of this type was proposed by Holman,  Ramond and Ross \cite{Holman:1984yj}. We will briefly describe it here, because it will be closely related to the models of string theory inflation to be discussed in this paper.
The  K\"ahler potential in this model is $K = \frac{\Phi\bar\Phi}{M_p^{2}}$, and the superpotential  is
\be
W = c\, (\Phi-\Phi_{0})^{2}\ ,
\ee
where $\Phi = (\varphi +i\alpha)/\sqrt 2$, and $\varphi$ and $\alpha$ are canonically normalized scalar fields. The potential has a minimum at $\alpha = 0$, $\varphi = \sqrt 2 \Phi_{0}$.

In what follows, we will work in units $M_{p} = 1$. For the particular case $\Phi_{0}  = 1$, the potential $V(\varphi)$ is shown by the thick blue solid line in Fig. 1. It has a minimum at $\varphi = \sqrt 2$, and a small plateau at $|\varphi| \ll 1$, where the potential is described by the simple equation
\be\label{cubic}
V(\varphi) =V_{0}\, (1-\sqrt 2\, \varphi^{3} + O(1)\,\varphi^{4}) \ ,
\ee 
where $V_{0} = c^{2}$.

\begin{figure}
\centering{\includegraphics[height=5cm]{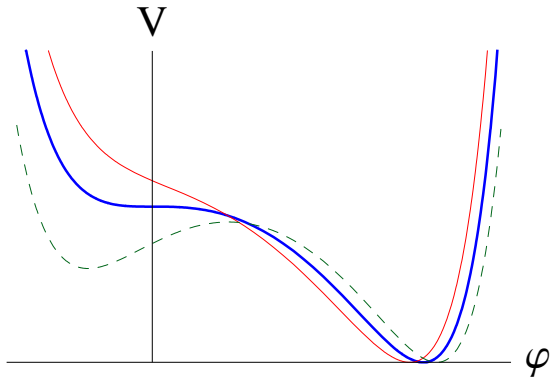}}
 \caption{The thick blue solid line shows the inflationary potential for the model of Ref. \cite{Holman:1984yj}, with $\Phi_{0} = 1$.  If one takes the model with a slightly larger $\Phi_{0}$, the potential has two minima separated by a barrier, as shown by the thin green dashed line, corresponding to $\Phi_{0} = 1.05$. For  $\Phi_{0}<1$, the metastable minimum disappears, and $V'$ becomes negative everywhere, as shown by the thin red solid line corresponding to $\Phi_{0} = 0.95$. Inflation is possible only if $|\Phi_{0}-1|\ll 1$, which requires fine tuning.} \label{Ramond}
\end{figure}

This potential is flat   at $\varphi = 0$, $V' = V'' = 0$, which corresponds to an inflection point. Since the slow-roll parameters $\epsilon$ and $\eta $ vanish at $\varphi = 0$, this potential can support inflation if the initial value of the field is sufficiently small. The slow-roll conditions in this model are satisfied for $\varphi \lesssim 10^{{-1}}$. The field $\varphi_{60}$ corresponding to the beginning of the period of the last 60 e-foldings is even much smaller. Therefore it is quite sufficient to use just  the first two terms of the expression (\ref{cubic}), i.e. the constant and the cubic term, for investigation of inflation in this model. 

On the other hand, there is no special reason why $\Phi_{0}$ must be exactly equal $1$. If we take, for example, $\Phi_{0}= 0.95$, we will obtain a steep potential shown by the thin red solid line in Fig. 1, and inflation disappears. For   $\Phi_{0}= 1.05$, the potential has two minima and a non-inflationary maximum with $V'' \sim V$. To achieve 60 e-folds of inflation in this scenario one must have $\Phi = 1$ with a great accuracy, and the universe from the very beginning should reside very close to the inflection point $\varphi = 0$. In this sense, inflation happens in the models of this type by accident.

This model was proposed  back in 1984. It took 10 years, from 1984 to 1994, until the hybrid inflation scenario was developed \cite{Hybrid} and it became possible to introduce the F-term \cite{F} and the D-term hybrid inflation scenario \cite{D}, where inflation could appear in a relatively natural way. It took 6 more years after that until the simple realization of  chaotic inflation with a quadratic potential was proposed in supergravity  \cite{Kawasaki:2000yn}, and it took 7 more years until a supergravity realization of natural inflation became possible \cite{Kallosh:2007ig}.

A systematic investigation of inflation in string theory began only in 2003, after the development of the KKLT mechanism of vacuum stabilization \cite{KKLT}, see also \cite{Giddings:2001yu,Silverstein:2001xn}. The first string inflation model based on the KKLT mechanism, the KKLMMT model  \cite{Kachru:2003sx}, is a brane inflation version of the hybrid inflation scenario. A detailed investigation of this model \cite{Baumann:2007np,delicate,Krause:2007jk,Panda:2007ie,Itzhaki:2007nk}
 demonstrated that its inflationary potential is very similar to the inflection point potential of Ref.  \cite{Holman:1984yj}.\footnote{Similar potentials emerge in other inflationary models as well, such as the MSSM inflation, see  Ref. \cite{Allahverdi:2006we}.}  Thus, in a certain sense, in the recent studies of string theory inflation we returned to the discussion of accidental inflation in the first models of inflation in supergravity.

In parallel with the investigation of the KKLMMT scenario, string theorists are trying to develop more natural versions of inflationary theory, where the inflationary potential would naturally have flat directions, such as D3/D7 inflation \cite{D3D7}. We do not know yet whether it is possible to obtain a generalization of the chaotic inflation scenario and/or of the natural inflation scenario; some related ideas are discussed e.g. in \cite{Dimopoulos:2005ac,Kallosh:2007ig,McAllister:2007bg,Grimm:2007hs}. As we just mentioned, the search for such models in supergravity took many years. The investigation of inflation in string theory has just began. Therefore at this early stage it may make a lot of sense to study the models of accidental inflation discussed above, despite the fact that they appear fine-tuned in more than one respect. This may be quite appropriate because the old ideas of naturalness, unnaturalness and fine-tuning look quite different in the context of the eternal inflation scenario in the string theory landscape consisting of $10^{10^{3}}$ different string theory vacua \cite{Lerche,Bousso,KKLT,Susskind:2003kw, Douglas:2003um}.

In this paper we will pursue two different goals. First of all,  the  dynamical mechanism of inflation in the KKLMMT model is quite complicated. It involves investigation of gravitational effects related to the motion of branes in warped space, which requires a consistent merger of open string theory and closed string theory. Therefore it would be nice to have a simple  toy model of accidental inflation in string theory, analogous to the model of Ref. \cite{Holman:1984yj}.  An example of such model will be presented in Section \ref{KLinf}.

Secondly, we would like to understand generic features of accidental inflation in the context of the string landscape scenario,  and, if possible, to find some unambiguous predictions of this class of models which would allow us to distinguish it from other versions of string inflation. These issues will be discussed in Sections  \ref{toyinf} -- \ref{predict}. We will summarize our results in Section \ref{concl}. A short discussion concerning eternal inflation can be found in the Appendix.

\section{Volume modulus inflation}\label{KLinf}

For a long time, it did not seem possible that the volume modulus may play the role of the inflaton field in the KKLT construction. Indeed, the standard KKLT potential has only one minimum at finite values of the volume modulus, and the slow-roll conditions for this potential are strongly violated~\cite{alphaprime}.

The situation did not change much with the invention of the racetrack inflation: It became possible to find a saddle point of the potential and satisfy the slow-roll conditions  with respect to the axion field, but not with respect to  the volume modulus \cite{Blanco-Pillado:2004ns}.

In this section, we propose a model of string theory inflation driven by the volume modulus in the context of the KL~\cite{Kallosh:2004yh} model of moduli stabilization. (Another model of the volume modulus inflation will be described soon  in \cite{CKLQ}.) Here the combination of fluxes with a racetrack superpotential leads to a scalar potential with two local asymmetric minima for the volume modulus which become de Sitter after uplifting by e.g. \d3bar-branes or \ds-induced D-terms. The higher of the two minima connects through a saddle point to the late-time dS minimum, or degenerates into an inflection point. With a proper choice of parameters, one can make the potential near the saddle point or the inflection point  sufficiently flat to provide for 
more than the necessary 60 e-foldings of slow-roll inflation driven by the volume modulus.

To begin with, we shall thus shortly review the structure of the KL setup~\cite{Kallosh:2004yh}. The model starts out similar to the KKLT construction~\cite{KKLT} by assuming that the complex structure moduli and the axion-dilaton have been frozen at a high KK-related mass scale by turning on sufficiently generic 3-form fluxes $G_3$ along the lines of~\cite{Giddings:2001yu}. With respect to the K\"ahler moduli this results in a constant and fine-tunable term $W_0=\int_{\rm CY_3}G_3\wedge \Omega$ in superpotential of the low-energy effective 4d ${\cal N}=1$ supergravity. Considering then the subclass of Calabi-Yau 3-folds with $h^{1,1}=1$, i.e. a single K\"ahler modulus which is the volume modulus $T$, the presence of non-perturbative effects leads to the stabilization of $T$. These exponential terms
arise  either from Euclidean \dt-branes of from gaugino condensation  on \ds-branes, as explained in
\cite{KKLT,Blanco-Pillado:2004ns}. If there is only one exponential term in the superpotential, then this leads directly to the KKLT construction.

The KL model is then specified by assuming the presence of non-perturbative contributions in $T$ to the superpotential, leading to a racetrack superpotential. Thus the setup is given by
\bea
K &=& - 3 \ln(T + \overline{T})\;,\qquad T=\varphi+i\tau \ , \nn\\ &&\\
W &=& W_0 + Ae^{-aT}+ Be^{-bT} \ ,\nn
\label{setup}
\eea
where the volume modulus $\varphi$ measuring the volume of the single 4-cycle of the Calabi-Yau pairs up with the RR-axion $\tau$ to form the complex chiral superfield $T$.

The F-term scalar potential, \be V= e^{K}\left(  G^{T \overline T} D_{T}W \overline {D_{T} W}
- 3|W|^2 \right), \ee as the function of the volume modulus $\varphi$ is given by
\bea
&&V= \frac{e^{-2(a+b)\varphi}}{6\varphi^2}(b B e^{a\varphi}+ a A e^{b\varphi})\nonumber
\\ &\times& \left [Be^{a\varphi}(3+b\varphi)+e^{b\varphi}(A(3+a\varphi)+3e^{a\varphi}W_0)\right].
\label{pot}\eea 
It can be shown that this is a minimum in the axion direction, provided we take we take  $A,a$, $B, b$ and $W_0$ to be  all real
and the sign of $A$ and $B$ opposite, as well as $a>b$ and $A>|B|$.

Up to this point, the KL model does not differ from other racetrack models. However, as it was found in \cite{Kallosh:2004yh},  the scalar potential in this model for some values of its parameters has  {\it two} supersymmetric AdS minima which solve $D_T W=0$. The one at smaller volume is given (similarly to the heterotic racetrack models) by
\be
 \varphi_{cr}= \frac{1}{a-b}\ln \left |\frac{a\,A}{b\,B}\right| \ ,
\label{ReTcr} \ee
while the one at larger volume is the deeper AdS minimum of the two with its position determined similarly to the KKLT case
\be
\varphi_{_{\rm KKLT-like}}\sim-\frac{1}{a}\ln(W_0/A)\;.\label{ReTAdS} 
\ee
In the special case where one arranges for a particular relation
between the  parameters of the superpotential,
\be -W_0= A \left |\frac{a\,A}{
b\,B}\right|^\frac{a}{b-a} +B \left |\frac{a\,A}{b\,B}\right| ^\frac{b}{b-a}\ , \ee 
the smaller volume `racetrack-like' minimum actually becomes a SUSY Minkowski one with vanishing gravitino mass~\cite{Kallosh:2004yh}.

\begin{figure}
\centering{\hskip 1cm ~ \includegraphics[height=5cm]{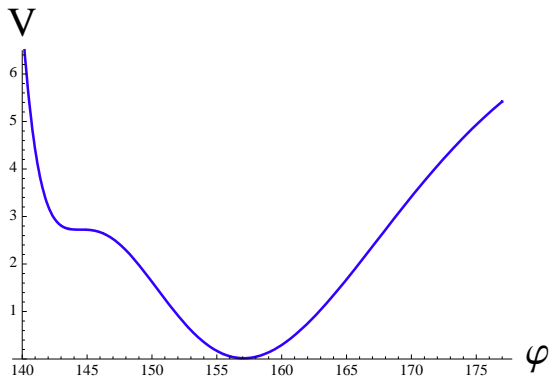}}
 \caption{The uplifted potential for $W_0= 4\cdot 10^{-8},\ A=1,\ B=-0.62704017319,\ a=2\pi/58,\
b=2\pi/60,\ C=3.01\cdot 10^{-18}$. The potential is shown in units of $10^{{-23}}$ of the Planck density. A late-time dS minimum at $V\approx 0$ stabilizes the volume at $\varphi_{\rm dS}\approx 157.1$. A tunable dS saddle or inflection point, which is responsible for inflation, is at $\varphi_{cr}\approx 144.5$.  There is a barrier at $\varphi \sim 200$ protecting the late-time dS minimum. The height of the barrier exceeds the height of the inflationary saddle/inflection point implying an absence of the cosmological overshoot problem~\cite{Brustein:1992nk,kaloper}.} \label{3}
\end{figure}

\begin{figure}
\centering{\includegraphics[height=6.5cm]{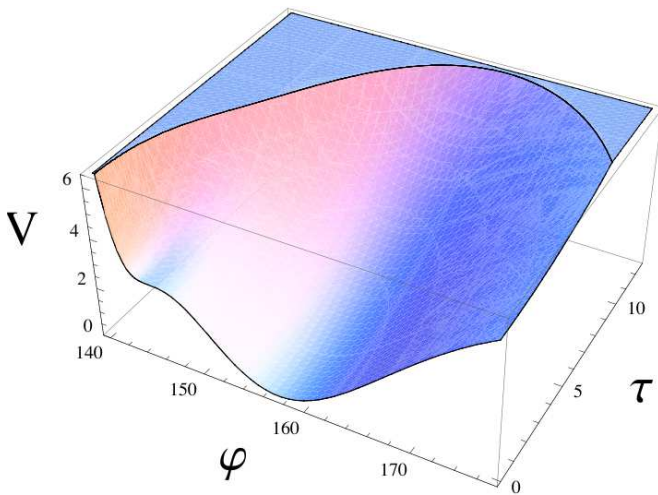}}
 \caption{The potential as a function of the complex field $T = \varphi +i\tau$,  in units of $10^{{-23}}$ of the Planck density. The inflationary saddle/inflection point and the late-time dS minimum both
occur at $\tau = {\rm Im}~T =0$, as shown in the analytic
investigation.} \label{4}
\end{figure}

One can then get a late-time de Sitter vacuum by adding, e.g. \d3bar-branes, which due to warping provide for a hierarchically small supersymmetry breaking and uplifting term $C/\varphi^2$ in the scalar potential.\footnote{Alternatives include D-terms from a 4-cycle wrapping \ds-brane threaded by $F_2$-flux~\cite{Dterms}, F-terms of hidden sector matter fields~\cite{nilles} (see also~\cite{ReinoScrucca}), meta-stable vacua of condensing gauge theories in the free magnetic range along the lines of ISS~\cite{ISSmodels}, and $\alpha'$-corrections~\cite{alphaprimelift}} This lifting can then be used to either lift the smaller volume (and more shallow) AdS/Minkowski minimum eq.~\eqref{ReTcr} to a tiny positive value of the cosmological constant, or to lift the deeper AdS minimum eq.~\eqref{ReTAdS}. In this latter possibility,  the smaller volume minimum eq.~\eqref{ReTcr} becomes much more strongly lifted and acquires a large positive cosmological constant. Since in many cases it will still be separated from the lower dS minimum at larger volume by a barrier, this specific sub-setup realizes the situation depicted in the context of the toy model of the last Section in Fig.~\ref{Ramond}.

It is now clear that by tuning $W_0$ and (through their dependence on the VEV's of the complex structure moduli) also $A$ and $B$ with fluxes, we can arrange for the higher-lying minimum at smaller volume to degenerate with the barrier saddle point. As this will guarantee the existence of a very flat saddle or inflection point, which can lead to inflationary regime. For instance, a choice of parameters
\bea W_0&=& 4\cdot 10^{-8},\ A=1,\ B=-0.62704017319,\nn \\ &&\\ a&=&\frac{2\pi}{58},\ b=\frac{2\pi}{60}\;{\rm and}\;C=3.01\cdot 10^{-18}\nn\label{input}\eea 
for the full uplifted scalar potential
\be
V= e^{K}\left(  G^{T \overline T} D_{T}W \overline {D_{T} W}
- 3|W|^2 \right)+\frac{C}{\varphi^2}
\ee
leads - as discussed below - to a long period of volume modulus inflation of more than 190 e-folds.

The similarity of this model with the supergravity model of Ref. \cite{Holman:1984yj} is obvious when one compares Fig.~1 and  Fig.~\ref{3}, while Fig.~\ref{4} demonstrates that $\tau=0$ indeed denotes a local minimum in the axion direction for all values of $\varphi$.

We shall now describe the inflationary dynamics for the parameters chosen in eq.~\eqref{input} in more detail. For these parameters we find that the minimum given by eq.~\eqref{ReTcr} has already merged completely with the barrier saddle point to form a so-called inflection point: Instead of the former saddle point that provides for $\epsilon=0$ by definition and needs than $\eta$ to be tuned small, we now have a situation where $\eta=0$ at the inflection point and we have to keep $\epsilon$ small.

The slow-roll parameters of the inflection point have now to be given for the case of non-canonically normalized scalar field, as the metric on scalar field space for $\varphi$ is given by $g_{\varphi\varphi}=3/2\varphi^2$~\cite{Blanco-Pillado:2004ns}. Since the inflaton trajectory is one-dimensional (as we are in the local minimum of the axion direction, there is no field evolution in this direction), we arrive for the chosen set of parameters at values~\cite{Blanco-Pillado:2004ns}
\bea \epsilon&=&\frac{g^{\varphi\varphi}}{2}\left(\frac{\der_\varphi V}{V}\right)^2=\frac{\varphi^2}{3}\left(\frac{\der_\varphi V}{V}\right)^2=1.8\cdot 10^{-18}\nn\\ && \\ \eta&=&g^{\varphi\varphi}\frac{\der_\varphi^2 V}{V}=\frac{2\varphi^2}{3}\frac{\der_\varphi^2 V}{V}\equiv 0\nn\eea
at the inflection point at $\varphi_{cr}=144.4588$.

\begin{figure}[ht!]
\centering\leavevmode\epsfysize=5.4cm \epsfbox{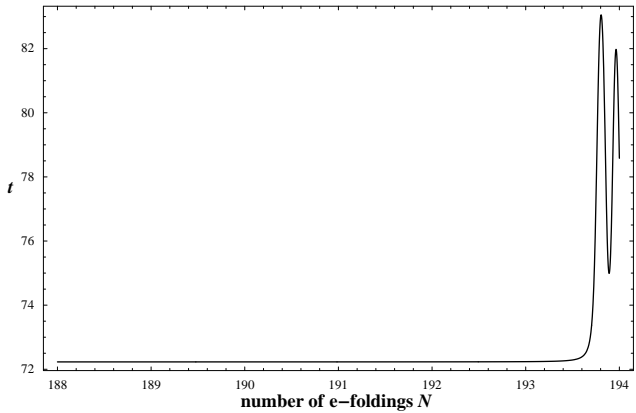} \caption{The evolution of the volume modulus $t$ as a function of the total number of e-folds $N$.} \label{efolds}
\end{figure}

The corresponding inflationary regime can be directly seen by numerically solving the equations of motion for $\varphi$. In transforming into measuring time by e-folds $N$
\be \frac{\der \varphi}{\der t}=H\frac{\der \varphi}{\der N}\equiv H\varphi' \ee
with
\bea H^2&=&\frac{1}{3}\left(\frac{1}{2}g_{\varphi\varphi} \dot\varphi^2+V(\varphi)\right)\nn\\ &=&\frac{1}{3}\frac{V(\varphi)}{1-\frac{1}{6}g_{\varphi\varphi} \varphi'^2}\eea we get the equation of motion to be~\cite{Blanco-Pillado:2004ns,noncanon}
\begin{eqnarray}
 \varphi''&=&-\left(1-\frac{1}{6}g_{\varphi\varphi} \varphi'^2\right)\left(3\varphi'+3g^{\varphi\varphi}\frac{V_{,\varphi}}{V}\right)\nonumber\\
 &-&\frac{\varphi'^2}{2}\frac{\der\ln(g_{\varphi\varphi})}{\der\varphi}\;.\label{eom} 
\end{eqnarray}
Here we have allowed for a non-canonical kinetic term ${\cal L}_{\rm kin}=\frac{1}{2}g_{\varphi\varphi}\der_\mu\varphi\der^{\mu}\varphi$, as this is the generic case when dealing with moduli.

According to eq.~\eqref{eom} and plugging in $g_{\varphi\varphi}$, the equation of motion for the volume modulus $\varphi$ then takes the form~\cite{Blanco-Pillado:2004ns}
\be
\varphi''=-\left(1-\frac{\varphi'^2}{4\varphi^2}\right)\left(3\varphi'+2\varphi^2\frac{V_{,\varphi}}{V}\right)+\frac{\varphi'^2}{\varphi}\;.\label{teom}
\ee
Starting at the inflection point $\varphi(t =0)=\varphi_0$ with $\dot \varphi(0)=0$, this leads to about 193 e-folds of slow-roll inflation, before the volume modulus rolls off into the late-time dS minimum at $\varphi_{\rm dS}\approx 157.1$. The CMB scales of COBE normalization at about $k_{\rm CMB}\simeq 0.002/{\rm Mpc}$  leave the horizon about 60 e-folds before the end of inflation, i.e. here at $N_{\rm CMB}\approx 133$. The magnitude of the primordial curvature perturbation $\Delta_{\cal R}^2$ generated at this point evaluates for the parameters chosen to be
\be \Delta_{\cal R}^2=\frac{1}{8\pi^2}\frac{H^4}{{\cal L}_{\rm kin}}=\frac{1}{4\pi^2}\frac{H^2}{g_{\varphi\varphi}\varphi'^2}\approx 2.9\cdot 10^{-9}\ee
which agrees with the measured value of COBE and WMAP~\cite{WMAP}.
In Fig.~\ref{efolds} we see the last 20 e-folds of the $t$-evolution. Note the sharp end of inflation at $N_{\rm tot}\approx 193.5$ and the subsequent onset of oscillations.

\begin{figure}[ht!]
\vspace*{0.5cm}
\centering\leavevmode\epsfysize=5.3cm \epsfbox{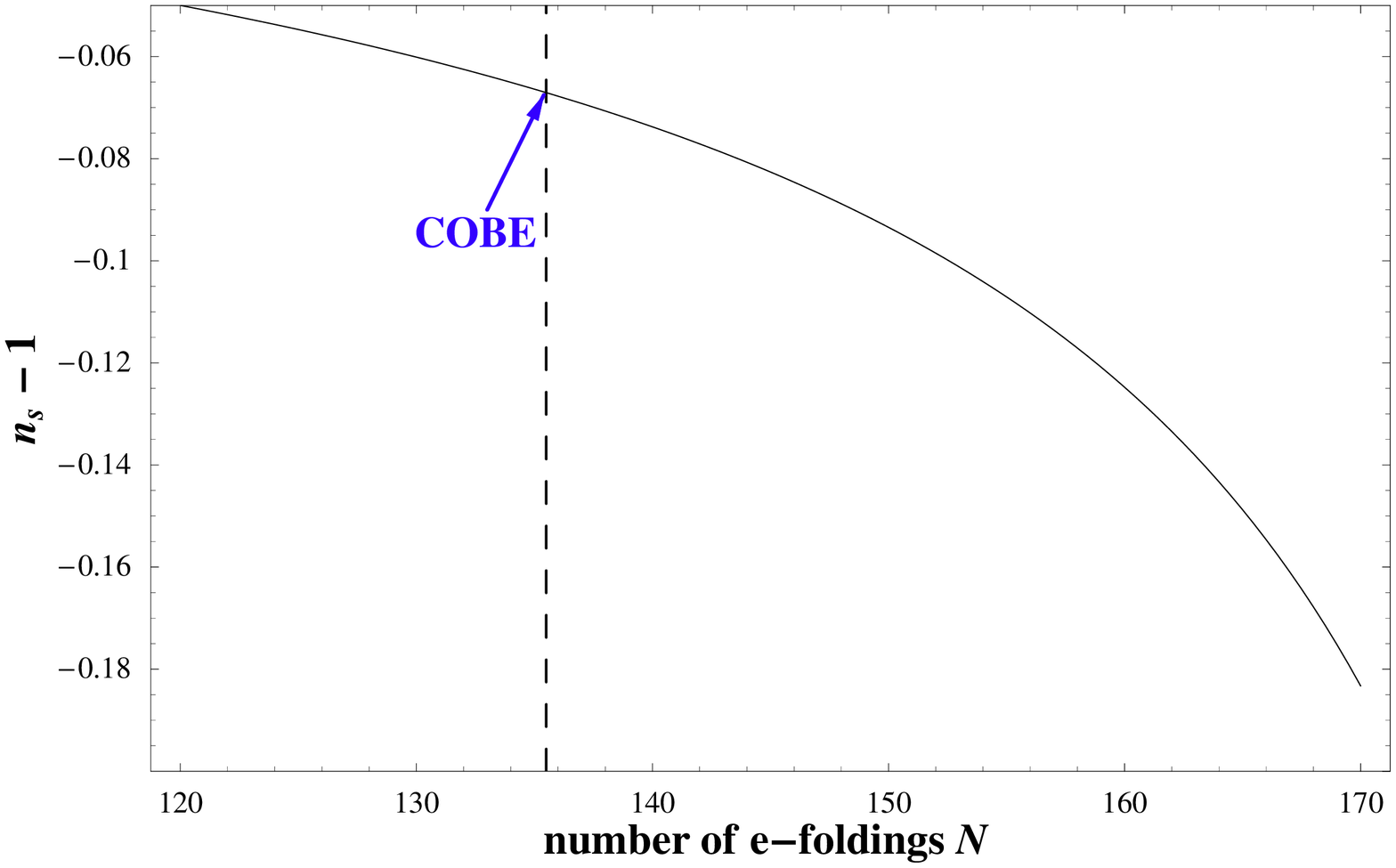} \caption{The spectral index of the density fluctuations as a function of the total number of e-folds $N$. The blue arrow denotes the spectral index at the time when the COBE normalization scale left the horizon, i.e. at $N_e^{\rm CMB}=60$ e-folds before the end of inflation.} \label{specindex}
\end{figure}

Next, the spectral index of the curvature perturbation power spectrum is given by \be
n_s=1+\left.\frac{d\ln \Delta_{\cal R}^2(k)}{d\ln
k}\,\right|_{k=RH}\label{specind2}\ee evaluated as usual at horizon crossing.
Note that here we can replace $d\ln k\simeq dN$ because $k$ is evaluated at
horizon crossing $k=RH\sim H e^N$. Then we arrive at \be n_s=1+\left.\frac{d\ln\Delta_{\cal R}^2(k)}{dN}\right|_{k=RH}\ee which results in the curve shown in
Fig.~\ref{specindex}.

This evaluates at the COBE normalization scale
\be n_s\approx 0.93\ee
Within the $1\sigma$-level this result agrees
with the combined 3-year WMAP + SDSS galaxy survey
result $n_s=0.948^{+0.015}_{-0.018}$~\cite{WMAP} (the 3-year WMAP
data alone give $n_s=0.951^{+0.015}_{-0.019}$).

We note further, that there is only negligible running at CMB scales as the third order slow-roll parameter evaluates to be $\xi^2\approx -7\cdot 10^{-4}$ which then gives $dn_s/d\ln k=2\xi^2\approx-0.0014$, and an extremely small level of tensor fluctuations as $\epsilon\approx 5 \cdot 10^{-17}$ at CMB scales.

\section{Basics of fine-tuning in generic models}\label{toyinf}

Now that we have two simple models of accidental inflation with an inflection point, the supergravity model of \cite{Holman:1984yj} and the string inflation model described in the previous section, and also the models of brane inflation  \cite{Baumann:2007np,delicate,Krause:2007jk,Panda:2007ie,Itzhaki:2007nk} and MSSM inflation  \cite{Allahverdi:2006we}, we can analyze general features of the models of this type.

Fine-tuning in the model \cite{Holman:1984yj} can be described in the following general way. If one takes $\Phi_{0}> 1$ in this model, one will have a potential with a metastable dS minimum at $\varphi < 0$ very close to the local maximum at $\varphi > 0$, see Fig. 1. If the difference $\Phi_{0}- 1$ is sufficiently small, then in the first approximation the term $V_{0}$ and the coefficient in front of $\varphi^{3}$ in (\ref{cubic}) will not change, but two other terms with very small coefficients $\lambda_{1},\lambda_{2} \ll 1$ will appear:
\be\label{cubic2}
V(\varphi) =V_{0}\, \bigl(1 + \lambda_{1}\varphi + \frac{\lambda_{2}}{ 2} \varphi^{2}-\sqrt 2\, \varphi^{3} + O(1)\,\varphi^{4}\bigr) \ .
\ee 
In the limit $\Phi_{0} \to 1$, the metastable minimum moves towards the local maximum, and they form an inflection point. This corresponds to the limit $\lambda_{1},\lambda_{2} \to 0$.  The slow-roll parameters at $\varphi = 0$ are directly related to $\lambda_{1}$ and $\lambda_{2}$:
\bea
\epsilon &=& \frac{1}{2}\left(\frac{V'}{V}\right)^{2} = \frac{1}{2} \lambda_{1}^{2} \ ,\nonumber\\
\eta &=& \frac{V''}{V} =    \lambda_{2} \ .
\eea

Thus in order to achieve inflation one should first find a potential with a metastable dS minimum close to a local maximum of $V(\varphi)$, and then change the parameters in such a way as to make these two extrema very close, which automatically will make the slow-roll parameters small. The required fine-tuning can be expressed in terms of the fine-tuning of the parameters $\lambda_{1}$ and $\lambda_{2}$.

One can look at this situation from a slightly different but closely related perspective. In accidental inflation, by definition, inflation occurs by chance, in a small vicinity of some point which can be called $\varphi =0$ without any loss of generality.
Consider the potential in a vicinity of this point and expand it in powers of $\varphi$:
\be\label{cubic1}
V(\varphi) =V_{0}\, \Bigl(1 + \lambda_{1}\varphi + \frac{\lambda_{2}}{ 2} \varphi^{2} + \frac{\lambda_{3}}{3}\, \varphi^{3} + \frac{\lambda_{4}}{4}\,\varphi^{4}+...\Bigr)
\ee 
If the potential has flat directions, as in the simplest versions of chaotic or natural inflation, then one may consider theories where only few first terms are present \cite{Linde:2007fr}. However, in most of the versions of string inflation all  parameters $\lambda_{n}$ can be large, so making some of them small requires fine-tuning. Slow roll inflation is possible for $|\varphi | \ll 1$ if $|\lambda_{1}| \ll 1$ and $|\lambda_{2}|\ll 1$. We will assume that $\lambda_{3} \gtrsim 1$, which is indeed the case in the two models which we considered so far. Unless one fine-tunes $\lambda_{4} \gg \lambda_{3}$, one can ignore the quartic term while describing inflation at  $\varphi \ll 1$. This means that  in accidental inflation, which occurs only at $\varphi \ll 1$, it is generically sufficient to use only three first terms in the expansion of the potential. 

Moreover, one can show that by a shift $\varphi \to \varphi + \Delta$ in a theory with $\lambda_{1},\lambda_{2} \ll 1$, $\lambda_{3} \gtrsim 1$, one can always represent the potential (\ref{cubic1}) in the form where either $\lambda_{1} = 0$, or $\lambda_{2} = 0$. For example, in the model of Ref. \cite{Holman:1984yj}  one has three possibilities illustrated by Fig. 1.
For $\Phi_{0} = 1$ one has an inflection point at $\varphi = 0$ where $\lambda_{1} = \lambda_{2} = 0$. For $\Phi_{0} > 1$ inflation begins near a local maximum of the potential at $\varphi > 0$. If one makes a change of variables $\varphi \to \varphi + \Delta$, where $\Delta$ corresponds to the position of this maximum, the potential acquires the form
\be\label{23}
V(\varphi) =V_{0}\, \Bigl(1 +  \frac{\lambda_{2}}{ 2} \varphi^{2} + \frac{\lambda_{3}}{3}\, \varphi^{3} + ...\Bigr) \ .
\ee
with $\lambda_{2} < 0$. Meanwhile for $\Phi_{0} > 1$ one can always make a change of variables bringing the potential to the form 
\be\label{13}
V(\varphi) =V_{0}\, \Bigl(1 +  {\lambda_{1} } \varphi  + \frac{\lambda_{3}}{3}\, \varphi^{3} + ...\Bigr) \ .
\ee

We ignored here the quartic term because  accidental inflation,  by definition, occurs only for $|\varphi| \ll 1$. Therefore, unless one makes an additional fine-tuning $\lambda_{3} \ll \lambda_{4}$, one can ignore the term $\frac{\lambda_{4}}{4}\,\varphi^{4}$ in the description of inflationary dynamics in these models.\footnote{We should emphasize that some of our conclusions are based on the assumption that in the class of the models we study now inflation occurs only for $|\varphi| \ll 1$, which is the trademark of accidental inflation. In other words, we study generic features of the worst case scenario. It should remain our goal to search for the models with large flat directions.}

An important exception is represented by the models where the potentials have a maximum with $Z_{2}$ symmetry $\varphi \to -\varphi$. In this case
\be\label{24}
V(\varphi) =V_{0}\, \Bigl(1 + \frac{\lambda_{2}}{ 2} \varphi^{2} + \frac{\lambda_{4}}{4}\,\varphi^{4}+...\Bigr)  \ ,
\ee 
and the required fine-tuning is $|\lambda_{2}|\ll 1$. In this case inflation is most efficient in the limit  $\lambda_{2} \to 0$, which corresponds to the purely quartic potential with $\lambda_{4} < 0$. This is what happens in the original versions of the new inflation scenario \cite{New}. This regime was also found in the racetrack inflation models \cite{Blanco-Pillado:2004ns}, where inflation occurs during the rolling in the axion direction from the saddle point of the potential.

Note that it is insufficient to simply take  $\lambda_{1}\ll 1$ or $\lambda_{2} \ll 1$, respectively, in our models. The required fine-tuning should be somewhat stronger as we want to achieve more than 50 or 60 e-folds of inflation. 

As an example, one may consider the potential \eqref{13} describing an inflection point  and check how small should be the constant $\lambda_{1}$. Investigation of this situation was performed in~\cite{delicate}, and result for the total number of e-folding during inflation is 
\be
N_{\rm tot}=\int_{0}^\infty \frac{ V d\varphi}{V'} = \int_{0}^\infty \frac{d\varphi}{\sqrt{2\epsilon}}=\frac{\pi}{2}\frac{1}{\sqrt{\lambda_1\lambda_3}} \ .\label{NtotBaum}
\ee
In order to avoid significant running of the spectrum, one must have $N_{\rm tot} \gg N_e^{\rm CMB}\simeq 60$. This leads to a condition
\be
\lambda_{1} \ll 10^{{-3}} \lambda_{3}^{{-1}} \lesssim 10^{-3} \ ,\label{tunerunning}
\ee
whereas we already assumed that $\lambda_{3} \gtrsim 1$ (this assumption is justified below). Thus we must have fine-tuning at least of the order $10^{-3}$.  We must fine-tune $\lambda_{2}$ in a similar way in the models (\ref{23}), (\ref{24}). 

While this fine-tuning  is certainly not nice, in the next section we will explain  possible reasons which may make it looking quite natural.

The values of the slow roll parameters, the duration of inflation,  and the spectral index $n_{s} $  depend only on the $\lambda_{n}$, but they do not depend on the overall factor $V_{0}$. Meanwhile the square of the amplitude of metric perturbations $\Delta_{\cal R}^2$ scales as $V_{0}$. This allows to achieve an inflationary regime with desirable properties by fine-tuning  $\lambda_{1}$ or $\lambda_2$, and then   dial the desirable amplitude of perturbations of metric by changing $V_{0}$ without changing $\lambda_{n}$ any further \cite{Blanco-Pillado:2004ns}.

\section{Classical rolling versus quantum diffusion}\label{diffusion}

The structure of the model discussed before -- inflation starts driven by a single field direction from either a flat saddle or a flat inflection point of the potential -- lends itself to an analytical treatment of the inflationary dynamics with respect to the properties of the power spectrum of density fluctuations generated during slow-roll.

The starting point is the observation, that due to the fact that the whole inflationary dynamics is governed by very small field displacements $\Delta \varphi\ll 1$ away from the saddle/inflection point at $\varphi_{\rm cr}$, we can canonically normalize the field $\varphi$ in the vicinity of the unstable point and describe its motion by expanding the potential around the unstable point and `renormalizing' every derivative with respect to $\varphi$ as $\der_\varphi \to \sqrt{g^{\varphi\varphi}(\varphi_{\rm cr})}\, \der_\varphi$. This leads to slow-roll parameters defined as
\bea \epsilon&=&\frac{\varphi_{\rm cr}^2}{3}\left(\frac{\der_\varphi V}{V}\right)^2\nn\\ && \\ \eta&=&\frac{2\varphi_{\rm cr}^2}{3}\frac{\der_\varphi^2 V}{V}\quad.\nn\eea
Next, we write the expansion eq.~\eqref{cubic1} of the potential around the unstable point as
\be\label{Vexp}
V(\varphi) =  V_0\left(1+\sum_{p\geq 1}\frac{\lambda_p}{p}\,\varphi^p\right) \ .
 \ee
Here $\lambda_p=(g^{\varphi\varphi})^{p/2}\, {V^{(p)}_0}/{V_0}$, \, 
$V_0^{(p)}=\der_\varphi^p V(\varphi_{\rm cr})$, {and we replace }\, $\varphi\to \varphi+\varphi_{\rm cr}$, 
 so that $\varphi=0$ now denotes the critical point and $\varphi$ denotes what was $\Delta\varphi$ before.

Depending on whether the potential is monotonic around the unstable point or not, we get either an expansion around an inflection point solving $\lambda_2\sim\der_\varphi^2 V(\varphi_{\rm cr})=0$ or a saddle point solving $\lambda_1\sim\der_\varphi V(\varphi_{\rm cr})=0$. Successful inflation requires us then to tune small either $\lambda_1$ (inflection point) or $\lambda_2$ (saddle point).

Following the discussion of the last Section, we can thus parametrize the leading approximation to the potential in the two cases as
\be 
V=\begin{cases}V_0\left(1-\lambda_1\varphi-\frac{\lambda_3}{3}\varphi^3\right)  & \text{for an inflection point}\\
V_0\big(1+\frac{\lambda_2}{2}\varphi^2\pm\frac{\lambda_p}{p}\varphi^p\big) & \text{for a saddle}
\end{cases}\ee
where we have
\be
\lambda_2=\eta_0\equiv\eta(0)=-m^2/V_0<0\ .
\ee

We subdivide the second case into saddle points which are subject to a symmetry $Z_2:\,\varphi\to-\varphi$ (then $p=4+2n,~n\in \mathbb{N}$ starts with 4) and saddle points which are not (then p starts with 3).

We can then compute within the validity of the slow-roll regime the number of e-foldings $N_e(\varphi)$ counting backwards from the value $N_e(\varphi_{\rm end})\equiv 0$ at the end of inflation at $\varphi_{\rm end}$
\be
N_e(\varphi)=\int_{\varphi_{\rm end}}^\varphi \frac{d\varphi}{\sqrt{2\epsilon}}\;.
\ee
This integral can be performed in closed form in both cases. From the expression for $N_e(\varphi)$ we can then express the slow-roll parameters $\eps$ and $\eta$ as function of $N_e$.

Since we need to fine-tune $\lambda_{1}, \lambda_{2}$ to be extremely small, we will start our investigation with the study of the simple model 
\be
V=V_0\left(1-\frac{\lambda_p}{p}\varphi^p\right) .\label{Vsimp}
\ee
In this simplified situation we have
\be
N_e(\varphi_N)=\frac{1}{p-2}\cdot\frac{1}{\lambda_p\varphi_N^{p-2}}\ , \label{NePbic}
\ee
where we have taken $\varphi_{\rm end}\to\infty$ as this does not change the result significantly. From this expression we find the slow-roll parameters to be
\bea\label{etansPbic}
\eps&=&\frac{1}{2(p-2)^{2\frac{p-1}{p-2}}}\lambda_p^{-\frac{2}{p-2}} N_e^{-2\frac{p-1}{p-2}}\nn\\ && \\ \eta&=&-\,\frac{p-1}{p-2}\cdot\frac{1}{N_e}\quad.\nn
\eea
We can then compute the density fluctuations at $N_e$ e-folds before the end of inflation
\bea\label{DensflucAnalyt}
\left.\Delta_{\cal R}^2\right|_{N_e}&=&\frac{1}{4\pi^2}\left(\frac{H^2}{\dot\varphi}\right)_{N_e}^2=\frac{1}{24\pi^2}\left.\frac{V}{\eps}\right|_{N_e}\ , \nn\\ &&
 \\
&=&\frac{V_0(p-2)^{2\frac{p-1}{p-2}}}{12\pi^2}\lambda_p^{\frac{2}{p-2}} N_e^{2\frac{p-1}{p-2}}\ .\nn\eea

As the quantity $\left.\Delta_{\cal R}^2\right|_{N_e}$ is constrained by the COBE and WMAP results at the scale of $N_e=N_e^{\rm CMB} \sim 60$, this equation leads to a relation between the scale of the potential $V_0$ and its small-field power-law behavior controlled by $\lambda_p$.

We shall illustrate this constraint for the case $p=3$ of the explicit string model of the Section~\ref{KLinf} where
\be
p=3\ ,\quad \lambda_3\approx 3.7\cdot 10^{4}\ ,\quad V_0\approx 2.7\cdot 10^{-23} \ , \label{ExampleVal}
\ee
which implies then from the eq.~\eqref{DensflucAnalyt} above that
\be
{V_0} \approx  3\cdot 10^{-14}\, \lambda_3^{-2} 
\ee
in units of Planck density. (In our paper we use Planck mass $M_{\rm P}=2.4\cdot 10^{18}\,{\rm GeV}$ and work in units $M_{p}= 1$.)

Upon inspection of eq.~\eqref{NePbic} we might conclude that the total number of e-folds is infinite if we were to start inflation at $\varphi\to0$. However, our classical equations ignored quantum diffusion of the field $\varphi$, which is important near the point $\varphi = 0$. 

Indeed, the field does not move classically from this point because $V' = 0$ there. However, the long-wavelength perturbations grow as follows:
\be
\langle \varphi^{2} \rangle  = \frac{H^{3}}{4\pi^{2}}t \ .
\ee
  These perturbations are stretched by inflation, and therefore for all practical purposes they look as a nearly homogeneous classical field with a typical magnitude 
\be
\varphi = \sqrt{\langle \varphi^{2} \rangle}  = \frac{H}{2\pi }\sqrt {Ht} \  
\ee  
and speed 
\be
\dot \varphi   = \frac{H}{4\pi }\sqrt \frac{H}{ t} = \frac{H^{3}}{8\pi^{2} \varphi} \  .
\ee  
The speed of the classical motion,  $\dot\varphi = V'/3H$,  becomes greater than the speed of diffusion for
\be
|\varphi\, V'| = V_{0 }\lambda_{p} \varphi^{p} > \frac{V_{0}^{2}}{24\pi^{2}} \ .
\ee
Thus  the field grows due to quantum diffusion until some moment $\bar t$ when its average value grows up to 
\be\label{phibar}
\bar\varphi  \sim \left(\frac{V_{0}}{24\pi^{2}\lambda_{p}}\right)^{1/p} \ .
\ee
During this part of the process the size of the universe grows by a factor of $e^{\bar N} \sim e^{H\bar t}$, where
\be
\bar N \sim \lambda_{p}^{-2/p}\, V_{0}^{2/p - 1}\frac{(24\pi^{2})^{1-2/p}}{2} \ .
\ee
After that, the universe continues growing in the classical regime. According to Eq. (\ref{NePbic}), the number of e-folds in this regime is given by
\be\label{expen}
N_e(\bar\varphi)\sim \lambda_{p}^{-2/p}\, V_{0}^{2/p - 1}\frac{(24\pi^{2})^{1-2/p}}{p-2}=\frac{2\bar N}{p-2} \ .
\ee
This gives
\be\label{expen2}
N_{\rm tot} = \bar N + N_{e}\sim \frac{1}{2}\lambda_{p}^{-2/p}\, V_{0}^{2/p - 1} (24\pi^{2})^{1-2/p}\ \frac{p}{p-2} \ .
\ee

Let us consider now a realistic situation with $p = 3$, $\lambda_{3} = 1$ and $V_{0} \sim 3\times 10^{{-14}}$. In this case the total growth of volume of the universe can be estimated by
\be
e^{3N_{\rm tot}} \sim e^{10^{6}} \ .
\ee
The analogous result for the quartic potential, p = 4, with $\lambda_{4} = O(1)$, is 
\be
e^{3N_{\rm tot}} \sim e^{3\cdot 10^{8}} \ .
\ee

Notice that the expression for $N_{\rm tot}$ still diverges for $\lambda_p\to0$. Clearly, this divergence is unphysical as for $\lambda_p\lesssim 1$ we would find that the potential reaches $V \approx 0$ only for $\varphi >1$ in Planck units. This would lead to chaotic inflation at large $\varphi$, which should be treated separately.

In this paper we consider  only situations where $\varphi\ll 1$, $\lambda_p\gtrsim 1$. In the explicit realization of the Section~\ref{KLinf} this always fulfilled by a large margin, $\lambda_p\gg 1$, once one expands the potential there around $t_{\rm cr}$ according to eq.~\eqref{Vexp}. Furthermore, demanding $\lambda_p\gtrsim 1$ implies that $\bar\varphi\sim V_0^{1/p}\ll 1$ as long as $V_0\ll 1$ which justifies the simplifications made before in that the inflationary process really takes place very close to the unstable point.

The constraint $\lambda_p\gtrsim 1$ plugged into eq.~\eqref{etansPbic} implies an upper bound on $\eps$ as now we have
\be
\eps \lesssim N_e^{-2\frac{p-1}{p-2}}\leq \frac{1}{N_e^4}\quad\text{for }p\geq 3\quad.
\ee
Further, from eq.~\eqref{etansPbic} we have a spectral index~\cite{LythRiotto,hilltop}
\be
n_s=1+2\eta=1-2\,\frac{p-1}{p-2}\cdot\frac{1}{N_{e}} \ .\label{nsAnalyt}
\ee

Observationally we probe $N_e=N_e^{\rm CMB}\simeq 60$. Thus, we infer an immeasurably low fraction $r=12.4\,\eps_{60}\lesssim 10^{-6}$ of power in gravitational waves and a spectral index which changes from $0.93$ for $p = 3$ to $0.97$ for $p \gg 3$.

We may now test our methods by applying them to the explicit string theory model proposed in Section \ref{KLinf}. Instead of calculating the amplitude of perturbations directly, we may represent the potential as a sum of terms $\lambda_{n}\varphi^{n}$ up to $n = 3$ using eq.~(\ref{Vexp}), and then use the methods developed in this section. Using the values of eq.~\eqref{ExampleVal} gives us in terms of eq.s.~\eqref{DensflucAnalyt} and \eqref{nsAnalyt}
\be
\Delta_{\cal R}^2\approx 4\cdot 10^{-9}\ ,
\quad n_s\approx 0.93\ ,
\ee
which are evaluated at the time of horizon crossing of the CMB normalization scale modes at $N_e\simeq 60$.
The agreement with the numerical results of  Section \ref{KLinf} is good enough to justify the use of the simplified potential, where we have neglected the small linear term and higher powers $p\geq 4$.

For completeness, we note that the analogous treatment of the important exception represented by a $Z_2$-symmetric saddle point (quartic potential with a very small quadratic piece) yields
\be
n_s^{(p=4)}=1-\frac{3}{N_e^{\rm CMB}}\approx 0.95\ .
\ee

Finally, we may give here the results for $N_{\rm tot}$ and $n_s$ for finite values of $\lambda_{1,2}$.

We will consider the case of a $p=3$ inflection point with finite $\lambda_1$ and a general $p\geq 3$ saddle point with finite $\lambda_2$. In this situation, $N_{\rm tot}$ and $n_s$ will be a function of $\lambda_1$ or $\lambda_2$, respectively, for fixed $\lambda_p,\; p\geq 3$. We get again 
\be
N_{\rm tot}\simeq N_e(\bar\varphi)\simeq \frac{\pi}{2}\frac{1}{\sqrt{\lambda_1\lambda_3}}\quad\text{if: }\lambda_1\gtrsim\lambda_3\bar\varphi^2 \sim 10^{-10}
\ee
for the $p=3$ inflection point and
\be\label{Ntotsaddlepoint}
N_{\rm tot}\simeq N_e(\bar\varphi)=\frac{\lambda_2^{-1}}{p-2}\ln(1-\lambda_2\lambda_p^{-1}\bar\varphi^{2-p})
\ee
for the saddle point. Using these expressions we can compute $\eta(N_{\rm tot})$ by expressing $\lambda_{1,2}$ through the appropriate $N_{\rm tot}$ and we get $n_s(N_{\rm tot})$. The results are~\cite{delicate,Aterm}
\bea
n_s&=&1+2\eta=1-\frac{2\pi}{N_{\rm tot}}\cot\left(\frac{\pi N_e}{2N_{\rm tot}}\right)\nn\\ &=&1-\frac{4}{N_e}\ ,\quad\text{for }N_{\rm tot}\gg N_e\label{nsinflec}
\eea
for the $p=3$ inflection point and
\bea\label{nssaddle}
n_s&=&1+2\eta=1-2\lambda_2\left(1+\frac{p-1}{e^{(p-2)\lambda_2 N_e}-1}\right)\nn\\ &=&1-2\frac{p-1}{p-2}\cdot\frac{1}{N_e }\ ,\\&&\nn \\
&&\quad\text{for}\; \ N_{\rm tot}\gg N_e\quad\Leftrightarrow\quad  |\lambda_2|=|\eta_0|\ll 1\;.\nn
\eea
for a saddle point~\cite{Aterm}, where $\lambda_2=\lambda_2(N_{\rm tot})$ is the inverse of eq.~\eqref{Ntotsaddlepoint}.

\begin{figure}[ht!]
\hspace*{-0.2cm}\leavevmode\epsfysize=5.6cm \epsfbox{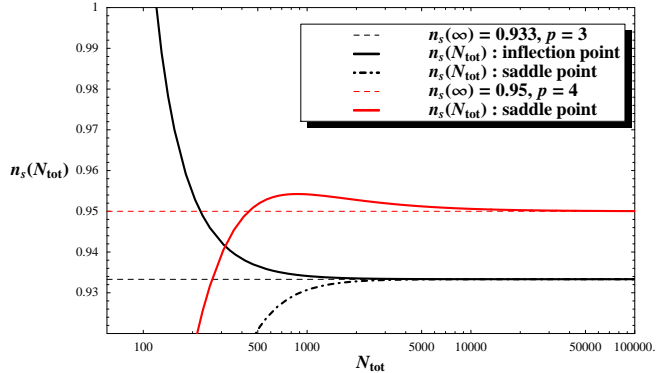} \caption{The spectral index $n_s$ as a function of the total number of e-folds $N_{\rm tot}$ at the COBE scale, i.e. in eq.s~\eqref{nsinflec} and \eqref{nssaddle} at $N_e=N_e^{\rm CMB}=60$. Black: The solid line corresponds to the $p=3$ inflection point. The dash-dotted line represents the $p=3$ saddle point case. The dashed horizontal line is their common asymptotic value $n_s=0.933$. Red: The solid line gives $n_s(N_{\rm tot})$ for the case $p=4$ with a $Z_2$-symmetry, as e.g. realized in the model of `racetrack inflation'~\cite{Blanco-Pillado:2004ns}. From there the values of $V_0$ and $\lambda_4$ have been extracted using eq.~\eqref{Vexp} analogously to the $p=3$ case treated here explicitly. This allows to compute $n_s(N_{\rm tot})$ in the same way. The dashed horizontal line gives the $p=4$ asymptotic value of $n_s=0.95$.} \label{specindex_analytic}
\end{figure}

We see that  we get back the results valid in the pure $\lambda_p\varphi^p$-potential for $N_{\rm tot}\gg N_e$ which corresponds to either very small $\lambda_1$ or $\lambda_2$. This behavior is displayed in Fig.~\ref{specindex_analytic} for the three cases of a $p=3$ inflection point (solid black), a $p=3$ saddle point (dash-dot black), and a $p=4$ $Z_2$-symmetric saddle point (solid red).

\section{Spectral index $n_{s}$ for accidental inflation}\label{predict}

\subsection{Naturalness of fine tuning}

We can now re-enter the linear or quadratic term, respectively, into the two cases of an inflection or a saddle point. Keeping them will enable to derive the dependence of spectral properties like $n_s$  on $\lambda_{1}$ and $\lambda_{2}$.

Let us see first what will happen if we add back to our cubic potential the term $\lambda_{1}\varphi$ with $\lambda_{1} < 0$. In this case the field $\varphi$ will roll down a bit faster, and the total number of e-folding will be smaller. The corresponding effect will be relatively insignificant if
\be\label{tuning}
|\lambda_{1}| \ll  |\lambda_{3}|\bar\varphi^{2} \sim (24\pi^{2})^{-2/3} \, |\lambda_{3}|^{1/3} V_{0}^{2/3}  \ .
\ee
For example, if one considers the semi-realistic case
$\lambda_{3} = {\cal O}(1)$ and $V_{0} \sim 3\cdot 10^{{-14}}$, this constraint means that for
\be\label{tuning2}
|\lambda_{1}| \ll  10^{{-10}} 
\ee
the average total growth of volume during inflation in  a vicinity of any given point will remain as large as $e^{3 \cdot 10^{5}}$. On the other hand, if we consider the case which, naively, would seem much more natural, $|\lambda_{1}|\gg 10^{-10}$, then the average total growth of volume during inflation will become exponentially smaller. As we already mentioned, the total degree of inflationary expansion will exceed the required 60 e-foldings for much smaller fine-tuning, $|\lambda_{1}| < 10^{{-3}}$, but it produces a growth of volume which is smaller than the volume produced for $|\lambda_{1}| \ll 10^{-10}$ by the enormous factor of $e^{-3 \cdot 10^{5}}$.

Now let us discuss the fine tuning of initial conditions in accidental inflation. According to (\ref{phibar}), if initially the absolute value of the field $\varphi$ in our part of the universe was smaller than  $\bar\varphi  \sim \left(\frac{V_{0}}{24\pi^{2}\lambda_{3}}\right)^{1/3} \sim 10^{-5}$, then the volume of this part grows  by an enormous factor $e^{10^{6}}$ (for $\lambda_{3} \sim 1,  V_{0} \sim 3\times 10^{{-14}}$ and $\lambda_{1} \sim 10^{{-10}}$). However, if the field initially was at a greater distance from the inflection point, the gain in volume immediately becomes much smaller. This is an important fact to keep in mind while discussing naturalness of initial conditions for inflation. An additional factor is the process of eternal inflation, which also occurs only for small $\varphi$, see Appendix.

An interesting feature of our results is that the total growth of volume does not increase when we fine-tune $\lambda_{1}$ with an accuracy greater than $10^{-10}$. In other words, our considerations do not predict $\lambda_{1} = 0$; they predict $\lambda_{1} \sim 10^{{-10}}$.

Similarly, if we add the quadratic term $\lambda_{2} \varphi^{2}/2$ to the model with $\lambda_{3} = O(1)$ and $V_{0} \sim 10^{{-14}}$, the total amount of inflation will be unaffected for $|\lambda_{2}| \ll  10^{{-5}}$,
and it will decrease dramatically for $|\lambda_{2}| \gg  10^{{-5}}$.

Finally if we consider the $Z_2$-symmetric class of models with a quartic potential with $\lambda_{4} = O(1)$, and add to it the quadratic term $\lambda_{2} \varphi^{2}/2$, then the average total growth of volume during inflation in  a vicinity of any given point remains in the range of $e^{8 \cdot 10^{7}}$ for $|\lambda_{2}| \ll  10^{{-10}}$, and it will rapidly drop down for $|\lambda_{2}| \gg  10^{{-10}}$.

Let us now discuss the consequences of these results for understanding of naturalness/unnaturalness of fine-tuning required for accidental inflation. Here we should remember that inflation in string theory landscape is always eternal because of the existence of an incredibly large number of metastable vacua \cite{Lerche,Bousso,KKLT,Susskind:2003kw, Douglas:2003um}. The problem of assigning probabilities for various histories in the eternal inflation scenario is still a subject of intense debates. One of the popular ideas is that the probability to be born in a particular part of the landscape is proportional to  the volume of the parts of the landscape of a given type, see e.g. \cite{LLM,Mediocr,Bellido2,Meas1,Linde:2006nw,LindeMeas,Hawking:2007vf} and references therein. Many authors agree that the volume-weighted probability should be proportional to the growth of volume during the slow-roll inflation \cite{Meas1,LindeMeas,Hawking:2007vf,LindeWinitz}.

If this is the case, then the results obtained above suggest that the fine-tuning of $\lambda_{1}$ and $\lambda_{2}$ required for accidental inflation is in fact quite natural; see \cite{Mediocr,Bellido2,Linde:2007fr} for a discussion of related ideas, and the Appendix of our paper where we reach similar conclusions using slightly different assumptions.

\subsection{A prediction for $n_s$}

These arguments also imply that the most probable values for the spectral index in this class of models will be the asymptotic values of eq.~\eqref{nsAnalyt} for the pure cubic (or quartic, in case of a $Z_2$-symmetric potential) case discussed above:

Using the volume-weighted measure, we should expect to find ourselves in a Hubble volume with a long period of slow-roll inflation in the past. And as we noted above, maximizing the amount slow-roll inflation implies having either $\lambda_1\to 0$ or $\lambda_2\to 0$ corresponding to a very large $N_{\rm tot}\gg N_e^{\rm CMB}$.

Looking again at Fig.~\ref{specindex_analytic}, it becomes clear that the asymptotic values of eq.~\eqref{nsAnalyt} for the pure cubic (or quartic, in case of a $Z_2$-symmetric potential)
\be
\begin{cases} n_s\approx 0.93\quad\text{no }Z_2\text{-symmetry of }V\\
n_s\approx 0.95\quad Z_2\text{-symmetry of }V\quad
\end{cases}
\ee
become a tentative prediction for accidental inflation in the string theory landscape. Together with the negligible tensor fraction $r<10^{-6}$ this is the hallmark of accidental inflation in string theory. We call this prediction tentative because it is based on certain assumptions about the probability measure. Moreover, it may change if we are allowed to fine-tune even further, by fine-tuning $\lambda_{3}$ to zero (which would lead to a greater value of $n_{s}$), changing $V_{0}$, the amplitude of density perturbations, etc., see the Appendix for a discussion of this issue.

\section{Conclusion}\label{concl}

We have shown that the generic presence of racetrack superpotentials together with fluxes leads to a realization of volume modulus inflation in type IIB string theory. The central observation there is that the multi-exponential superpotential generically leads to more than one local minimum. This can be used to tune the higher-lying of two adjacent minima to (nearly) degenerate with the separating barrier into a very flat saddle or inflection point, which then drives slow-roll inflation. Following these considerations, in Section \ref{KLinf} we constructed a model of volume modulus inflation, which is one of the simplest models of string theory inflation available now.

Inflation in a vicinity of an inflection or saddle point requires fine-tuning of the model parameters and of the initial conditions.  
However, this simple setup describes both the majority of known closed-string modular inflation models~\cite{Blanco-Pillado:2004ns,alphaprime}, as well as the recently improved KKLMMT model~\cite{Baumann:2007np,delicate,Krause:2007jk,Panda:2007ie,Itzhaki:2007nk}.\footnote{The \dt-\ds-brane inflationary scenario~\cite{D3D7} does not fall into this class as it leads quite naturally to hybrid inflation.} Thus, the dynamics of accidental inflation described in this paper may represent one of the most generic classes of inflationary models, covering a large part of the landscape. The required fine tuning may appear quite natural if one takes into account that it dramatically increases the total growth of volume during inflation, and may even make inflation eternal.

This fact  leads to a statistical prediction of the spectral index $n_s\approx 0.93$, or $n_s\approx 0.95$, depending on whether the potential is $Z_2$-symmetric or not. Furthermore, the value $n_s\approx 0.95$ for $Z_2$-symmetric potentials by means of this symmetry is an upper limit (see Fig.~\ref{specindex_analytic}) as long as we tune just the first even power in such a potential, i.e. $\lambda_2$. to zero (see also~\cite{Blanco-Pillado:2004ns,racetrns}). These results,  combined with a prediction of a negligibly small tensor to scalar ratio  $r<10^{-6}$,   become a testable (tentative) prediction for a large class of models of inflation in string theory.
\vskip 0.3cm

\noindent {\bf Acknowledgments}: It is a pleasure to thank R. Bousso, J. Conlon, R. Kallosh, F. Quevedo, V. Vanchurin, A.~Vilenkin and  S.~Winitzki for useful discussions. This work was supported in part by NSF grant PHY-0244728 and by the Alexander-von-Humboldt Foundation.

\appendix*
\section{Quantum diffusion vs. eternal inflation}

\subsection{Expansion after eternal inflation regime}

Here we would like to take the opportunity to compare the picture of eternal inflation with the local view of an observer emerging from the quantum diffusion process on the top of the potential into the slow-roll phase described in Section~\ref{diffusion}.

If  quantum jumps of the field $\varphi$  dominate its classical rolling during a typical time  $H^{{-1}}$, then each domain of a size $H^{{-1}}$ will eternally split into many new domains, in some of which  the field will over and again jump back arbitrarily close to $\varphi=0$, forever re-starting the slow-roll process in different $H^{-1}$-sized domains. This leads to  eternal inflation~\cite{Vilenkin:1983xq,Linde:1986fd}. 
It occurs for fields satisfying the following generic condition \cite{Linde:1986fd}:
\bea
\langle\delta\varphi\rangle_{\rm quant}&\gtrsim&\delta\varphi_{\rm class.}(\Delta t=H^{-1})=-\,\frac{V'}{3H^2}(\varphi_*)\nn\\ &\Rightarrow&\; V^{3} \gtrsim 12 \pi^{2}(V')^{2} \ .
\eea
For the simplified potential eq.~\eqref{Vsimp}, eternal inflation occurs for $\varphi < \varphi_{*}$, where
\be
\varphi_*\sim  (12\pi^{2})^{-\frac{1}{2p-2}}\lambda_p^{-\frac{1}{p-1}}V_0^{\frac{1}{2p-2}}\ .
\ee

The probability measure in eternal inflation regime is somewhat ambiguous because we must compare infinities, which produces regularization-dependent results. Several different approaches to this question suggest that the probability should be proportional to the growth of volume during inflation in the slow-roll regime. We used a version of this prescription in the main text. A slightly different prescription suggests that we should 
be interested only in the total number of e-foldings after the end of eternal inflation.

In this case, we should take $\varphi\sim\varphi_*$ as the initial condition for the phase of slow-roll inflation, which  leads to a finite amount of slow-roll inflation
\be 
N^*_{\rm tot}=\int_{\varphi_{\rm end}}^{\varphi_*} \frac{d\varphi}{\sqrt{2\eps}} \sim  \frac{(12\pi^{2})^{\frac{p-2}{2p-2}}}{p-2}\, \lambda_p^{-\frac{1}{p-1}} V_0^{-\frac{p-2}{2p-2}}\quad.\label{NtotAnalyt}
\ee

In comparison with the boundary of quantum diffusion, $\bar\varphi$ in eq.~\eqref{phibar}, one finds that for a purely quadratic potential at the top one gets $\bar\varphi\sim\varphi_*$~\cite{LindeWinitz} while for all other $p\geq 3$ one typically has $\bar\varphi\ll\varphi_*$.

The main point, however, is that the condition eq.~\eqref{tuning}, which should be satisfied in order to get the maximum amount of slow-roll inflation after the end of either eternal inflation or quantum diffusion (i.e. $|\lambda_1|\ll |\lambda_p|\varphi_*^{p-1}$ or $|\lambda_1|\ll |\lambda_p|\bar\varphi^{p-1}$ for quantum diffusion or eternal inflation, respectively), for eternal inflation yields a condition
\be
|\lambda_1|\ll (12\pi^{2})^{-1/2}\, V_0^{1/2}  \ .
\ee
In particular, for $p = 3$, $\lambda_{3} = 1$,  $V_0\sim 3\cdot 10^{-14}$ we get the condition $|\lambda_1|\ll  10^{{-8}}$. As before, for such a small $\lambda_{1}$ we have an extremely large $N_{\rm tot}$, and therefore we have the same prediction,  $n_{s} \approx 0.93$. Similarly for the theory with $p = 4$ we find $n_{s} \approx 0.95$. Thus our main conclusions coincide with the conclusions obtained in this paper by a somewhat different method.

\subsection{Slow roll without eternal inflation}

All uncertainties with the choice of the volume weighted probability measure discussed  appear only for  $
|\lambda_1|\lesssim \lambda_1^*\equiv (12\pi^{2})^{-1/2}\, V_0^{1/2}$, when inflation is eternal and one deals with the problem of comparing infinities.
For $
|\lambda_1|\gtrsim \lambda_1^*\equiv (12\pi^{2})^{-1/2}\, V_0^{1/2}$ inflation is no longer eternal, and the results of the calculations of the total volume of the universe created by the stage of inflation become quite unambiguous. In this regime one can use the result of Ref. \cite{delicate}, $N_{\rm tot}=\frac{\pi}{2}\frac{1}{\sqrt{\lambda_1\lambda_3}} $, see Eq. (\ref{NtotBaum}).

This boundary value $\lambda_1^*\sim 10^{-8}$ is much smaller than the value $\lambda_1^{N_e>60}\sim 10^{-3}/\lambda_3$ needed for just 60 e-folds of slow-roll according to eq.~\eqref{tunerunning}. Thus, we see that for $\lambda_3\gtrsim 1$ there is a whole range of values $10^{-8}\lesssim \lambda_1\lesssim 10^{-3}$ where there occurs no eternal inflation at all, but a pro-longed phase of slow-roll inflation with $N_{\rm tot}\gg 60$. This result shows, unambiguously, that the total number of observers in any part of the universe after inflation is proportional to 
\be
e^{3N_{tot}} \sim \exp\left (\frac{3\pi}{2\sqrt{\lambda_{1}\lambda_{3}}}\right) \ .
\ee
This quantity acquires its maximal value at the smallest possible value of $\lambda_{1}$ within its domain of validity, i.e. at 
\be
\lambda_{1} \sim  \lambda_1^*\equiv (12\pi^{2})^{-1/2}\, V_0^{1/2} \sim 10^{{-8}}
\ee
 for $V_{0} \sim 3\times 10^{{-14}}$. Note that this constraint depends only on $V_{0}$ and is valid for any $p > 2$ independently of $\lambda_{p}$.

For $\lambda < \lambda_1^*$ eternal inflation becomes possible, which makes the total number of observers living in the post-inflationary universe indefinitely large. This shows that, independently of the choice of the probability measure during eternal inflation, the volume-weighted probability should be peaked at $\lambda \lesssim \lambda_1^* \sim 10^{{-8}}$. Once again, this leads to the prediction $n_{s} \approx 0.93$ in this class of models, in agreement with the predictions obtained by other methods.

A similar result can be derived for the saddle point case, where again for $\lambda_2\lesssim \lambda_2^*\equiv (V_0/12\pi^2)^{(p-2)/2(p-1)}\lambda_p^{1/(p-1)}$ the total number of slow-roll e-folds becomes independent of $\lambda_2$. In the above case this value is $\lambda_2^*\sim 10^{-4}$, which implies that there is a prolonged phase of slow-roll inflation with $N_{\rm tot}\gg 60$. Once again, this is an unambiguous prediction of the volume-weighted probability distribution, independently of the choice of the probability measure during eternal inflation in this scenario. This again leads to the prediction $n_{s} \approx 0.95$ for inflation near the symmetric saddle point with $p = 4$.

Of course, predictions of that type are only as good as the basic assumption that the probability to be born in parts of the universe of a certain type is proportional to the total number of observers in such parts. Whereas this assumption seems quite natural and its various versions are widely accepted \cite{LLM,Mediocr,Bellido2,Meas1,Linde:2006nw,LindeMeas,Hawking:2007vf}, there are some alternative proposals, which de-emphasize the growth of volume during inflation, see e.g.  \cite{susskind,bousso}.  These proposals will lead to different predictions for $n_{s}$, which opens a possibility to test various ideas concerning the probabilities in inflationary cosmology by comparing the consequences of these ideas with  observational data.

\subsection{Even more tuning? -- Maybe not}

It is rather interesting that the final result for $N_{\rm tot}$ in eq.~\eqref{expen2} can be represented in a form directly related to the amplitude of perturbations of metric:
\be\label{expepm}
N_{\rm tot} \sim \Delta_{\cal R}^{-2+4/p}\  C_{p} \ ,
\ee
where $C_{p} = p\,  \bigl(2(p-2)\bigr)^{-2/p}\, N_{e}^{2-2/p}$.
Note that dependence of the duration of inflation  on $V_{0}$ and $\lambda_{p}$ was totally absorbed in its dependence on $$ \Delta_{\cal R}^{2} \sim V_{0}\ \lambda_{p}^\frac{2}{p-2},$$ and the total growth of volume during inflation is given by
\be\label{durationpert}
e^{3N_{\rm tot}} \sim \exp\left({3\Delta_{\cal R}^{-2+4/p}\  C_{p}}\right) \ .
\ee
Taking this expressions to its limit would suggest that we might increase the volume factor further by decreasing the magnitude of metric perturbations which in turn corresponds to sending $V_0\to 0$ or $\lambda_p\to 0$. As it stands this would then predict that a universe like ours with a small but finite amount of metric fluctuations is extremely unlikely compared to universes with exponentially low fluctuation levels. This is the 'run-away' problem described  in~\cite{Mediocr,Bellido2,Garriga:2005ee,runaway2}.

Various solutions to this problem have been proposed in the literature, see e.g. \cite{Mediocr,Bellido2,Garriga:2005ee,Linde:2005yw}. Here we will specifically mention a  solution proposed in~\cite{runawaycure}. The essential point is that the prior probability distribution should besides the volume factor of slow-roll inflation also contain an anthropic constraint arising from the necessity of sufficiently fast radiative cooling of primordial baryonic matter (otherwise no structure forms before the protons decay) and the threshold behavior of inflation-related cosmological quantities in terms of the observable fundamental inflationary parameters. In many simple models the cosmological quantity is the reheating temperature $T_{\rm RH}\sim\sqrt{m_\varphi}$ being a function of the inflation mass~\cite{runawaycure}.

Firstly, the anthropic constraint from sufficiently fast cooling generates an exponential  suppression which counters the exponentially large volume factor. Secondly then, the threshold behavior of $T_{\rm RH}$ as a function of the inflation mass (the fundamental parameter) leads to the volume factor $\exp(3N_{\rm tot})$ being nearly flat when expressed as a function of $T_{\rm RH}$ in the threshold region of inflation decay (while, of course, it still depends exponentially on the inflaton mass). The combined probability distribution then is (unlike the volume factor alone) approximately flat in the region $10^{-12}<\Delta_{\cal R}^2<10^{-8}$ which allows then the anthropic argument for $\Delta_{\cal R}^2\sim 10^{-10}$ to proceed~\cite{runawaycure}.

One way or another, we should separate the issue of {\it explaining} the observed value of $\Delta_{\cal R}^2$ and {\it predicting} the value of $n_{s}$, which is not yet known with a sufficient precision, but will be known in the next few years; see \cite{Linde:2006nw,Garriga:2007wz} for a discussion of related issues.
It is in this sense  that we call the above results $n_s\approx 0.93$ or $n_s\approx 0.95$ tentative predictions for accidental inflation in the string theory landscape.


\begin{thebibliography}{99}

\bibitem{Chaot} A.~D.~Linde,
``Chaotic Inflation,'' Phys.\ Lett.\ B {\bf 129}, 177 (1983).

\bibitem{Linde:1985ub}
  A.~D.~Linde,
``Initial Conditions For Inflation,''
  Phys.\ Lett.\  B {\bf 162}, 281 (1985).
  
\bibitem{Vilenkin:1983xq}
  A.~Vilenkin,
``The Birth Of Inflationary Universes,''
  Phys.\ Rev.\  D {\bf 27}, 2848 (1983).

\bibitem{Linde:1986fd}
A.~D.~Linde,
``Eternal Chaotic Inflation,''
  Mod.\ Phys.\ Lett.\  A {\bf 1}, 81 (1986);
  A.~D.~Linde,
``Eternally Existing Self-reproducing Chaotic Inflationary Universe,''
  Phys.\ Lett.\  B {\bf 175}, 395 (1986).
  
\bibitem{Linde:2007fr}
  A.~Linde,
``Inflationary Cosmology,''
  arXiv:0705.0164 [hep-th].
  
\bibitem{Freese:1990rb}
  K.~Freese, J.~A.~Frieman \& A.~V.~Olinto,
``Natural inflation with pseudo - Nambu-Goldstone bosons,''
  Phys.\ Rev.\ Lett.\  {\bf 65}, 3233 (1990);
  F.~C.~Adams, J.~R.~Bond, K.~Freese, J.~A.~Frieman \& A.~V.~Olinto,
``Natural Inflation: Particle Physics Models, Power Law Spectra For Large
Scale Structure, And Constraints From COBE,''
  Phys.\ Rev.\  D {\bf 47}, 426 (1993)
  [arXiv:hep-ph/9207245].

\bibitem{Holman:1984yj}
  R.~Holman, P.~Ramond \& G.~G.~Ross,
``Supersymmetric Inflationary Cosmology,''
  Phys.\ Lett.\  B {\bf 137}, 343 (1984).
  

  
   
\bibitem{Hybrid}
A.~D.~Linde, ``Axions in inflationary cosmology,'' Phys.\ Lett.\ B
{\bf 259}, 38 (1991); A.~D.~Linde, ``Hybrid inflation,'' Phys.\
Rev.\ D {\bf 49}, 748 (1994) [astro-ph/9307002].

\bibitem{F}
E.~J.~Copeland, A.~R.~Liddle, D.~H.~Lyth, E.~D.~Stewart \&
D.~Wands, ``False vacuum inflation with Einstein gravity,'' Phys.\
Rev.\ D {\bf 49}, 6410 (1994) [astro-ph/9401011];
 G.~R.~Dvali,
Q.~Shafi \& R.~Schaefer, ``Large scale structure and
supersymmetric inflation without fine tuning,'' Phys.\ Rev.\
Lett.\  {\bf 73}, 1886 (1994) [hep-ph/9406319];
 A.~D.~Linde \& A.~Riotto,
``Hybrid inflation in supergravity,'' Phys.\ Rev.\ D {\bf 56},
1841 (1997) [arXiv:hep-ph/9703209].



\bibitem{D}
P.~Binetruy \& G.~Dvali, ``D-term inflation,'' Phys.\ Lett.\ B
{\bf 388}, 241 (1996) [hep-ph/9606342];
E.~Halyo, ``Hybrid
inflation from supergravity D-terms,'' Phys.\ Lett.\ B {\bf 387},
43 (1996) [hep-ph/9606423].

  
  
\bibitem{Kawasaki:2000yn}
  M.~Kawasaki, M.~Yamaguchi \& T.~Yanagida,
``Natural chaotic inflation in supergravity,''
  Phys.\ Rev.\ Lett.\  {\bf 85}, 3572 (2000)
  [arXiv:hep-ph/0004243].
    
  
\bibitem{Kallosh:2007ig}
  R.~Kallosh,
``On Inflation in String Theory,''
  arXiv:hep-th/0702059;
  R.~Kallosh, N.~Sivanandam \& M.~Soroush,
``Axion Inflation and Gravity Waves in String Theory,''
  arXiv:0710.3429 [hep-th].




    
    
    
  
\bibitem{KKLT}
  S.~Kachru, R.~Kallosh, A.~Linde \& S.~P.~Trivedi,
``De Sitter vacua in string theory,''
  Phys.\ Rev.\  D {\bf 68}, 046005 (2003)
  [arXiv:hep-th/0301240].
  
\bibitem{Giddings:2001yu}
  S.~B.~Giddings, S.~Kachru \& J.~Polchinski,
``Hierarchies from fluxes in string compactifications,''
  Phys.\ Rev.\  D {\bf 66}, 106006 (2002)
  [arXiv:hep-th/0105097].
  
\bibitem{Silverstein:2001xn}
  E.~Silverstein,
``(A)dS backgrounds from asymmetric orientifolds,''
  arXiv:hep-th/0106209;
  A.~Maloney, E.~Silverstein \& A.~Strominger,
``De Sitter space in noncritical string theory,''
  arXiv:hep-th/0205316.
  

\bibitem{Kachru:2003sx}
  S.~Kachru, R.~Kallosh, A.~Linde, J.~M.~Maldacena, L.~P.~McAllister \& S.~P.~Trivedi,
``Towards inflation in string theory,''
  JCAP {\bf 0310}, 013 (2003)
  [arXiv:hep-th/0308055].



\bibitem{Baumann:2007np}
  D.~Baumann, A.~Dymarsky, I.~R.~Klebanov, L.~McAllister \& P.~J.~Steinhardt,
``A Delicate Universe,''
  Phys.\ Rev.\ Lett.\  {\bf 99}, 141601 (2007)
  [arXiv:0705.3837 [hep-th]].


\bibitem{delicate}
D.~Baumann, A.~Dymarsky, I.~R.~Klebanov \& L.~McAllister,
``Towards an Explicit Model of D-brane Inflation,''
arXiv:0706.0360 [hep-th].

\bibitem{Krause:2007jk}
  A.~Krause \& E.~Pajer,
``Chasing Brane Inflation in String-Theory,''
  arXiv:0705.4682 [hep-th].
  
\bibitem{Panda:2007ie}
  S.~Panda, M.~Sami \& S.~Tsujikawa,
``Prospects of inflation in delicate D-brane cosmology,''
  arXiv:0707.2848 [hep-th].
  
\bibitem{Itzhaki:2007nk}
  N.~Itzhaki \& E.~D.~Kovetz,
``Inflection Point Inflation and Time Dependent Potentials in String
Theory,''
  arXiv:0708.2798 [hep-th].
  
\bibitem{Allahverdi:2006we}
  R.~Allahverdi, K.~Enqvist, J.~Garcia-Bellido, A.~Jokinen \& A.~Mazumdar,
``MSSM flat direction inflation: slow roll, stability, fine tuning and
reheating,''
  JCAP {\bf 0706}, 019 (2007)
  [arXiv:hep-ph/0610134].
  

\bibitem{D3D7}
R.~Kallosh, ``N = 2 supersymmetry and de Sitter space,''
arXiv:hep-th/0109168;
C.~Herdeiro, S.~Hirano \& R.~Kallosh,
``String theory and hybrid inflation/acceleration,'' JHEP {\bf
0112} (2001) 027 [arXiv:hep-th/0110271];
K.~Dasgupta, C.~Herdeiro,
S.~Hirano \& R.~Kallosh, ``D3/D7 inflationary model and
M-theory,'' Phys.\ Rev.\ D {\bf 65}, 126002 (2002)
[arXiv:hep-th/0203019];
J.~P.~Hsu, R.~Kallosh \& S.~Prokushkin, ``On brane inflation with
volume stabilization,'' JCAP {\bf 0312}, 009 (2003)
[arXiv:hep-th/0311077];
J.~P.~Hsu \& R.~Kallosh,
``Volume stabilization and the origin of the inflaton shift symmetry in string theory,''
  JHEP {\bf 0404}, 042 (2004)
  [arXiv:hep-th/0402047];
  K.~Dasgupta, J.~P.~Hsu, R.~Kallosh, A.~Linde \& M.~Zagermann,
``D3/D7 brane inflation and semi-local strings,''
  JHEP {\bf 0408}, 030 (2004)
  [arXiv:hep-th/0405247].
  
\bibitem{Dimopoulos:2005ac}
  S.~Dimopoulos, S.~Kachru, J.~McGreevy \& J.~G.~Wacker,
``N-flation,''
  arXiv:hep-th/0507205.

  
\bibitem{McAllister:2007bg}
  L.~McAllister \& E.~Silverstein,
``String Cosmology: A Review,''
  arXiv:0710.2951 [hep-th].
  
\bibitem{Grimm:2007hs}
T.~W.~Grimm,
``Axion Inflation in Type II String Theory,''
[arXiv:0710.3883 [hep-th]].



\bibitem{Lerche} W.~Lerche, D.~Lust \& A.~N.~Schellekens,
 ``Chiral Four-Dimensional Heterotic Strings From Self-dual Lattices,''
Nucl. Phys. B {\bf 287}, 477 (1987).


\bibitem{Bousso} R.~Bousso \& J.~Polchinski,
``Quantization of four-form fluxes and dynamical neutralization of the
cosmological constant,''
JHEP {\bf 0006}, 006 (2000)
[arXiv:hep-th/0004134].



\bibitem{Susskind:2003kw}
L.~Susskind,
``The anthropic landscape of string theory,''
arXiv:hep-th/0302219;

\bibitem{Douglas:2003um}
M.~R.~Douglas,
``The statistics of string / M theory vacua,''
JHEP {\bf 0305}, 046 (2003)
[arXiv:hep-th/0303194]
M.~R.~Douglas, B.~Shiffman \& S.~Zelditch,
``Critical points and supersymmetric vacua,''
arXiv:math.cv/0402326.
F.~Denef \& M.~R.~Douglas,
``Distributions of flux vacua,''
JHEP {\bf 0405}, 072 (2004)
[arXiv:hep-th/0404116];
A.~Giryavets, S.~Kachru \& P.~K.~Tripathy,
``On the taxonomy of flux vacua,''
JHEP {\bf 0408}, 002 (2004)
[arXiv:hep-th/0404243];
F.~Denef, M.~R.~Douglas \& B.~Florea,
``Building a better racetrack,''
JHEP {\bf 0406}, 034 (2004)
[arXiv:hep-th/0404257].

\bibitem{alphaprime}
A.~Westphal,
``Eternal inflation with alpha' corrections,''
JCAP {\bf 0511}, 003 (2005)
[arXiv:hep-th/0507079].


\bibitem{Blanco-Pillado:2004ns}
J.J. Blanco-Pillado, C.P. Burgess, J.M. Cline, C. Escoda, M. Gomez-Reino, R. Kallosh, A. Linde , F. Quevedo,
``Racetrack inflation,''
JHEP {\bf 0411}, 063 (2004)
[arXiv:hep-th/0406230];
Z.~Lalak, G.~G.~Ross \& S.~Sarkar,
``Racetrack inflation and assisted moduli stabilisation,''
  Nucl.\ Phys.\  B {\bf 766}, 1 (2007)
  [arXiv:hep-th/0503178];
J.J. Blanco-Pillado, C.P. Burgess, J.M. Cline, C. Escoda, M. Gomez-Reino, R. Kallosh, A. Linde , F. Quevedo,
``Inflating in a better racetrack,''
  JHEP {\bf 0609}, 002 (2006)
  [arXiv:hep-th/0603129];
  P.~Brax, A.~C.~Davis, S.~C.~Davis, R.~Jeannerot \& M.~Postma,
``D-term Uplifted Racetrack Inflation,''
[arXiv:0710.4876 [hep-th]].


\bibitem{Kallosh:2004yh}
R.~Kallosh \& A.~Linde,
``Landscape, the scale of SUSY breaking, and inflation,"
JHEP {\bf 0412}, 004 (2004)
[arXiv:hep-th/0411011];
J.~J.~Blanco-Pillado, R.~Kallosh \& A.~Linde,
``Supersymmetry and stability of flux vacua,''
  JHEP {\bf 0605}, 053 (2006)
  [arXiv:hep-th/0511042];
  A.~Ceresole, G.~Dall'Agata, A.~Giryavets, R.~Kallosh \& A.~Linde,
``Domain walls, near-BPS bubbles, and probabilities in the landscape,''
  Phys.\ Rev.\  D {\bf 74}, 086010 (2006)
  [arXiv:hep-th/0605266].


\bibitem{CKLQ} J.~P.~Conlon, R. Kallosh, A. Linde \& F. Quevedo, ``Volume modulus inflation and the gravitino mass problem,'' in preparation.



\bibitem{Dterms}
C.~P.~Burgess, R.~Kallosh \& F.~Quevedo,
``de Sitter string vacua from supersymmetric D-terms,''
JHEP {\bf 0310}, 056
(2003) [arXiv:hep-th/0309187];
A.~Achucarro, B.~de Carlos, J.~A.~Casas \& L.~Doplicher,
``de Sitter vacua from uplifting D-terms in effective supergravities from
realistic strings,''
[arXiv:hep-th/0601190];
  S.~L.~Parameswaran \& A.~Westphal,
  ``de Sitter string vacua from perturbative Kahler corrections and
  consistent D-terms,''
  JHEP {\bf 0610}, 079 (2006)
  [arXiv:hep-th/0602253].
E.~Dudas \& Y.~Mambrini,
``Moduli stabilization with positive vacuum energy,''
JHEP {\bf 0610}, 044 (2006)
[arXiv:hep-th/0607077];
M.~Haack, D.~Krefl, D.~Lust, A.~Van Proeyen \& M.~Zagermann,
``Gaugino condensates and D-terms from D7-branes,''
[arXiv:hep-th/0609211];
D.~Cremades, M.~P.~G.~del Moral, F.~Quevedo \& K.~Suruliz,
``Moduli stabilisation and de Sitter string vacua from magnetised D7
branes,''
[arXiv:hep-th/0701154].


\bibitem{nilles}
O.~Lebedev, H.~P.~Nilles \& M.~Ratz,
``de Sitter vacua from matter superpotentials,''
Phys.\ Lett.\  B {\bf 636}, 126 (2006)
[arXiv:hep-th/0603047].

\bibitem{ReinoScrucca}
M.~Gomez-Reino \& C.~A.~Scrucca,
``Locally stable non-supersymmetric Minkowski vacua in supergravity,''
JHEP {\bf 0605}, 015 (2006)
[arXiv:hep-th/0602246].


\bibitem{ISSmodels}
K.~Intriligator, N.~Seiberg \& D.~Shih,
``Dynamical SUSY breaking in meta-stable vacua,''
JHEP {\bf 0604}, 021
(2006) [arXiv:hep-th/0602239];
E.~Dudas, C.~Papineau \& S.~Pokorski,
``Moduli stabilization and uplifting with dynamically generated F-terms,''
[arXiv:hep-th/0610297];
H.~Abe, T.~Higaki, T.~Kobayashi \& Y.~Omura,
``Moduli stabilization, F-term uplifting and soft supersymmetry breaking
terms,''
Phys.\ Rev. D {\bf 75}, 025019 (2007)
[arXiv:hep-th/0611024]
F.~Br\"ummer, A.~Hebecker \& M.~Trapletti,
`SUSY breaking mediation by throat fields,''
Nucl.\ Phys.\ B {\bf 755}, 186 (2006) [arXiv:hep-th/0605232];
R.~Kallosh \& A.~Linde,
``O'KKLT"
[arXiv:hep-th/0611183];
M.~Serone \& A.~Westphal,
``Moduli Stabilization in Meta-Stable Heterotic Supergravity Vacua,''
JHEP {\bf 0708}, 080 (2007)
[arXiv:0707.0497 [hep-th]].


\bibitem{alphaprimelift}
V.~Balasubramanian \& P.~Berglund,
``Stringy corrections to Kahler potentials, SUSY breaking, and the
cosmological constant problem,''
JHEP {\bf 0411}, 085 (2004)
[arXiv:hep-th/0408054];
A.~Westphal,
``de Sitter String Vacua from Kahler Uplifting,''
JHEP {\bf 0703}, 102 (2007) [arXiv:hep-th/0611332];
S.~S.~AbdusSalam, J.~P.~Conlon, F.~Quevedo \& K.~Suruliz,
``Scanning the Landscape of Flux Compactifications: Vacuum Structure and Soft
Supersymmetry Breaking,''
arXiv:0709.0221 [hep-th].




\bibitem{Brustein:1992nk}
R.~Brustein \& P.~J.~Steinhardt,
``Challenges for superstring cosmology,''
Phys.\ Lett.\ B {\bf 302}, 196 (1993) [arXiv:hep-th/9212049].

\bibitem{kaloper} N.~Kaloper and K.~A.~Olive,
``Dilatons in string cosmology,'' Astropart.\ Phys.\  {\bf 1}, 185 (1993);T.~Barreiro, B.~de Carlos \& E.~J.~Copeland,
``Stabilizing the dilaton in superstring cosmology,'' Phys.\ Rev.\ D {\bf 58},
083513 (1998) [arXiv:hep-th/9805005];
A.~Linde,
``Creation of a compact topologically nontrivial inflationary universe,''
JCAP {\bf 0410}, 004 (2004)
[arXiv:hep-th/0408164];
R.~Brustein, S.~P.~de Alwis \& P.~Martens,
``Cosmological stabilization of moduli with steep potentials,''
Phys.\ Rev.\ D {\bf 70}, 126012 (2004)
[arXiv:hep-th/0408160];
N.~Kaloper, J.~Rahmfeld and L.~Sorbo,
``Moduli entrapment with primordial black holes,''
arXiv:hep-th/0409226.



\bibitem{noncanon}
R.~Kallosh, A.~Linde, S.~Prokushkin \& M.~Shmakova,
``Supergravity, dark energy and the fate of the universe,''
Phys.\ Rev.\  D {\bf 66}, 123503 (2002)
[arXiv:hep-th/0208156];
C.~P.~Burgess, P.~Grenier \& D.~Hoover,
``Quintessentially flat scalar potentials,''
JCAP {\bf 0403}, 008 (2004)
[arXiv:hep-ph/0308252];
C.~P.~Burgess, J.~M.~Cline, H.~Stoica \& F.~Quevedo,
``Inflation in realistic D-brane models,''
JHEP {\bf 0409}, 033 (2004)
[arXiv:hep-th/0403119].


\bibitem{WMAP}
D.~N.~Spergel {\it et al.},
``Wilkinson Microwave Anisotropy Probe (WMAP) three year results:
Implications for cosmology,''
Astrophys.\ J.\ Suppl.\  {\bf 170}, 377 (2007)
[arXiv:astro-ph/0603449].



\bibitem{New}
A.~D.~Linde, ``A New Inflationary Universe Scenario: A Possible
Solution Of The Horizon, Flatness, Homogeneity, Isotropy And
Primordial Monopole Problems,'' Phys.\ Lett.\ B {\bf 108}, 389
(1982); A.~Albrecht \& P.~J.~Steinhardt, ``Cosmology For Grand
Unified Theories With Radiatively Induced Symmetry Breaking,''
Phys.\ Rev.\ Lett.\  {\bf 48}, 1220 (1982); A.~D.~Linde,
``Coleman-Weinberg Theory And A New Inflationary Universe Scenario,''
  Phys.\ Lett.\ B {\bf 114}, 431 (1982);
A.~D.~Linde,
``Temperature Dependence Of Coupling Constants And The Phase Transition In
The Coleman-Weinberg Theory,''
  Phys.\ Lett.\ B {\bf 116}, 340 (1982);
A.~D.~Linde,
``Scalar Field Fluctuations In Expanding Universe And The New Inflationary
Universe Scenario,''
  Phys.\ Lett.\ B {\bf 116}, 335 (1982). 


\bibitem{LythRiotto}
D.~H.~Lyth \& A.~Riotto,
``Particle physics models of inflation and the cosmological density
perturbation,''
Phys.\ Rept.\  {\bf 314}, 1 (1999)
[arXiv:hep-ph/9807278].

\bibitem{hilltop}
L.~Boubekeur \& D.~H.~Lyth,
``Hilltop inflation,''
JCAP {\bf 0507}, 010 (2005)
[arXiv:hep-ph/0502047];
K.~Kohri, C.~M.~Lin \& D.~H.~Lyth,
``More hilltop inflation models,''
[arXiv:0707.3826 [hep-ph]].


\bibitem{Aterm}
  J.~C.~Bueno Sanchez, K.~Dimopoulos \& D.~H.~Lyth,
``A-term inflation and the MSSM,''
  JCAP {\bf 0701}, 015 (2007)
  [arXiv:hep-ph/0608299].



\bibitem{LLM}
  A.~D.~Linde \& A.~Mezhlumian,
``Stationary universe,''
  Phys.\ Lett.\  B {\bf 307}, 25 (1993)
  [arXiv:gr-qc/9304015];
A.~D.~Linde, D.~A.~Linde \& A.~Mezhlumian,
``From the Big Bang theory to the theory of a stationary universe,''
  Phys.\ Rev.\  D {\bf 49}, 1783 (1994)
  [arXiv:gr-qc/9306035];
J.~Garcia-Bellido, A.~D.~Linde \& D.~A.~Linde,
``Fluctuations Of The Gravitational Constant In The Inflationary Brans-Dicke
Cosmology,''
  Phys.\ Rev.\  D {\bf 50}, 730 (1994)
  [arXiv:astro-ph/9312039].

  
\bibitem{Mediocr}
  A.~Vilenkin,
``Predictions from quantum cosmology,''
  Phys.\ Rev.\ Lett.\  {\bf 74}, 846 (1995)
  [arXiv:gr-qc/9406010].
  
  \bibitem{Bellido2}
   J.~Garcia-Bellido \& A.~D.~Linde,
``Stationarity of inflation and predictions of quantum cosmology,''
  Phys.\ Rev.\  D {\bf 51}, 429 (1995)
  [arXiv:hep-th/9408023].

\bibitem{Meas1}
A.~Vilenkin, ``Making predictions in eternally inflating universe,''
Phys.\ Rev.\ D {\bf 52}, 3365 (1995)
[arXiv:gr-qc/9505031];
S.~Winitzki \& A.~Vilenkin,
``Uncertainties of predictions in models of eternal inflation,''
Phys.\ Rev.\ D {\bf 53}, 4298 (1996)
[arXiv:gr-qc/9510054];
J.~Garriga, D.~Schwartz-Perlov, A.~Vilenkin \& S.~Winitzki,
``Probabilities in the inflationary multiverse,''
JCAP {\bf 0601}, 017 (2006)
[arXiv:hep-th/0509184];
S.~Winitzki, ``Predictions in eternal inflation,''
[arXiv:gr-qc/0612164].
  

\bibitem{Linde:2006nw}
  A.~Linde,
  ``Sinks in the Landscape, Boltzmann Brains, and the Cosmological Constant
  Problem,''
  JCAP {\bf 0701}, 022 (2007)
  [arXiv:hep-th/0611043].
    
\bibitem{LindeMeas} A.~Linde,
``Towards a gauge invariant volume-weighted probability measure for   eternal 
inflation,''
JCAP {\bf 0706}, 017 (2007)
[arXiv:0705.1160 [hep-th]].

\bibitem{Hawking:2007vf}
S.~W.~Hawking,
``Volume Weighting in the No Boundary Proposal,''
[arXiv:0710.2029 [hep-th]];
J.~B.~Hartle, S.~W.~Hawking \& T.~Hertog,
``The No-Boundary Measure of the Universe,''
[arXiv:0711.4630 [hep-th]].

\bibitem{LindeWinitz}
A.~Linde, V.~Vanchurin \& S.~Winitzki, work in progress.
 
\bibitem{racetrns}
  Ph.~Brax, S.~C.~Davis \& M.~Postma,
  ``The Robustness of $n_s < 0.95$ in Racetrack Inflation,''
  [arXiv:0712.0535 [hep-th]].
 
\bibitem{susskind}
B.~Freivogel, M.~Kleban, M.~Rodriguez Martinez \& L.~Susskind,
``Observational consequences of a landscape,''
  JHEP {\bf 0603}, 039 (2006)
  [arXiv:hep-th/0505232].

\bibitem{bousso}
R.~Bousso,
``Holographic probabilities in eternal inflation,''
Phys.\ Rev.\ Lett.\  {\bf 97}, 191302 (2006)
[arXiv:hep-th/0605263];
R.~Bousso, B.~Freivogel \& I.~S.~Yang,
``Eternal inflation: The inside story,''
Phys.\ Rev.\  D {\bf 74}, 103516 (2006)
[arXiv:hep-th/0606114].

\bibitem{Garriga:2005ee}
  J.~Garriga \& A.~Vilenkin,
``Anthropic prediction for Lambda and the Q catastrophe,''
  Prog.\ Theor.\ Phys.\ Suppl.\  {\bf 163}, 245 (2006)
  [arXiv:hep-th/0508005].


\bibitem{runaway2}
  B.~Feldstein, L.~J.~Hall \& T.~Watari,
  ``Density perturbations and the cosmological constant from inflationary
  landscapes,''
  Phys.\ Rev.\  D {\bf 72}, 123506 (2005)
  [arXiv:hep-th/0506235].

\bibitem{Linde:2005yw}
  A.~Linde \& V.~Mukhanov,
  ``The curvaton web,''
  JCAP {\bf 0604}, 009 (2006)
  [arXiv:astro-ph/0511736].

\bibitem{runawaycure}
  L.~J.~Hall, T.~Watari \& T.~T.~Yanagida,
  ``Taming the runaway problem of inflationary landscapes,''
  Phys.\ Rev.\  D {\bf 73}, 103502 (2006)
  [arXiv:hep-th/0601028].




\bibitem{Garriga:2007wz}
  J.~Garriga \& A.~Vilenkin,
``Prediction and explanation in the multiverse,''
  arXiv:0711.2559 [hep-th].

\end{thebibliography}
\end{document}